\documentclass[epj]{svjour}

\usepackage{graphics}
\usepackage{times}

\begin{document}

\title{\hfill {\small FZJ-IKP-TH-2006-42}\\Hyperon-nucleon interactions in effective field theory}
\author{H. Polinder\inst{}\thanks{\email{h.polinder@fz-juelich.de}}}

\institute{Institut f{\"u}r Kernphysik (Theorie), Forschungszentrum J{\"u}lich, D-52425 J{\"u}lich, Germany}

\date{Received: date / Revised version: date}

\abstract{
We have constructed the leading order hyperon-nucleon potential in a chiral Effective Field Theory approach. The chiral potential consists of one-pseudoscalar-meson exchanges and non-derivative four-baryon contact terms. The hyperon-nucleon interactions are derived using SU(3) symmetry, the nucleon-nucleon interaction is not considered explicitly since it can not be described well with a leading order chiral Effective Field Theory. We solve a regularized Lippmann-Schwinger equation and show that a good description of the available hyperon-nucleon scattering data is possible with five free parameters. The chiral potential can be used as further input for hypernucleus and hypernuclear matter calculations. Preliminary hypertriton calculations yielded the correct hypertriton binding energy.
}
\maketitle

\section{Introduction}
\label{intro}
Since the work of Weinberg \cite{Wei90,Wei91}, the derivation of nuclear interactions from chiral Effective Field Theory (EFT) has been discussed extensively in the literature. For reviews we refer to \cite{Bed02,Epelbaum:2005pn}. The main advantages of this scheme are the possibilities to derive two- and three-nucleon forces as well as external current operators in a consistent way and to improve calculations systematically by going to higher orders in the power counting.

Recently the nucleon-nucleon ($NN$) interaction has been described to a high precision using chiral EFT \cite{Entem:2003ft,Epe05}. In \cite{Epe05}, the power counting is applied to the $NN$ potential, as originally proposed in \cite{Wei90,Wei91}. The $NN$ potential consists of pion-exchanges and a series of contact interactions with an increasing number of derivatives to parameterize the shorter ranged part of the $NN$ force. A regularized Lippmann-Schwinger equation is solved to calculate observable quantities. Note that in contrast to the original Weinberg scheme, the effective potential is made explicitely energy-independent as it is important for applications in few-nucleon systems (for details, see \cite{Epe98}).

The hyperon-nucleon ($YN$) interaction has not been investigated using EFT as extensively as the $NN$ interaction. Hyperon and nucleon mass shifts in nuclear matter, using chiral perturbation theory, have been studied in \cite{Sav96}. These authors used a chiral interaction containing four-baryon contact terms and pseudoscalar-meson exchanges. Recently, the hypertriton and $\Lambda d$ scattering were investigated in the framework of an EFT with contact interactions \cite{Ham02}. Some aspects of strong $\Lambda N$ scattering in effective field theory and its relation to various formulations of lattice QCD are discussed in  \cite{Beane:2003yx}. Korpa et al. \cite{Kor01} performed a next-to-leading order (NLO) EFT analysis of $YN$ scattering and hyperon mass shifts in nuclear matter. Their tree-level amplitude contains four-baryon contact terms; pseudoscalar-meson exchanges were not considered explicitly, but SU(3) breaking by meson masses was modeled by incorporating dimension two terms coming from one-pion exchange. The full scattering amplitude was calculated using the Kaplan-Savage-Wise resummation scheme \cite{Kap98}. The $YN$ scattering data were described successfully for laboratory momenta below 200 MeV, using 12 free parameters.

In this contribution we show the results for the recently constructed chiral EFT for the $YN$ system \cite{Pol06}. In this work we apply the scheme used in \cite{Epe05} to the $YN$ interaction. Analogous to the $NN$ potential, at leading order (LO) in the power counting, the $YN$ potential consists of pseudoscalar-meson (Goldstone boson) exchanges and four-baryon contact terms, related via SU(3) symmetry. We solve a regularized coupled channels Lippmann-Schwinger equation for the LO $YN$ potential and fit to the low-energy $YN$ cross sections, which are dominated by $S$-waves. We remark that our approach is quite different from \cite{Kor01}.

\section{The effective potential}
\label{sec:2}
In this section, we discuss in some detail the effective chiral $YN$ potential at leading order in the (modified) Weinberg power counting. This power counting is briefly recalled first. Then, we present the minimal set of non-derivative four-baryon interactions and show the formulae for the one-Goldstone-boson-exchange contributions.

\subsection{Power counting}
\label{sec:2.1}

We apply the power counting to the effective hyperon-nucleon potential $V_{\rm eff}$ which is then injected into a regularized Lippmann-Schwinger equation to generate the bound and scattering states. The various terms in the effective potential are ordered according to
\begin{equation}
\label{eq:1}
V_{\rm eff} \equiv V_{\rm eff}(Q,g,\mu) = \sum_\nu Q^\nu \, {\mathcal V}_\nu (Q/\mu ,g)~,
\end{equation}
where $Q$ is the soft scale (either a baryon three-momentum, a Goldstone boson four-momentum  or a Goldstone boson mass), $g$ is a generic symbol for the pertinent low-energy constants, $\mu$ a regularization scale, ${\mathcal V}_\nu$ is a function of order one, and $\nu \ge 0$ is the chiral power. It can be expressed as 
\begin{eqnarray}\label{eq:2}
\nu &=& 2 - B + 2L + \sum_i v_i \,\Delta_i ~,\nonumber\\
\Delta_i &=& d_i + {\displaystyle\frac{1}{2}}\, b_i  - 2~,
\end{eqnarray}
with $B$ the number of incoming (outgoing) baryon fields, $L$ counts the number of Goldstone boson loops, and $v_i$ is the number of vertices with dimension $\Delta_i$. The vertex dimension is expressed in terms of derivatives (or Goldstone boson masses) $d_i$  and the number of internal baryon fields $b_i$ at the vertex under consideration. The LO potential is given by $\nu = 0$, with $B=2$, $L=0$ and $\Delta_i = 0$. Using eq.~(\ref{eq:2}) it is easy to see that this condition is fulfilled for two types of interactions -- a) non-derivative four-baryon contact terms with $b_i = 4$ and $d_i = 0$ and b) one-meson exchange diagrams with the leading meson-baryon derivative vertices allowed by chiral symmetry ($b_i = 2,d_i = 1$). At LO, the effective potential is entirely given by these two types of contributions, which will be discussed in more detail in the following sections.

\subsection{The four-baryon contact terms}
\label{sec:2.2}
\begin{figure}[t]
\begin{center}
\resizebox{0.5\textwidth}{!}{\includegraphics*[2cm,21.0cm][18cm,26cm]{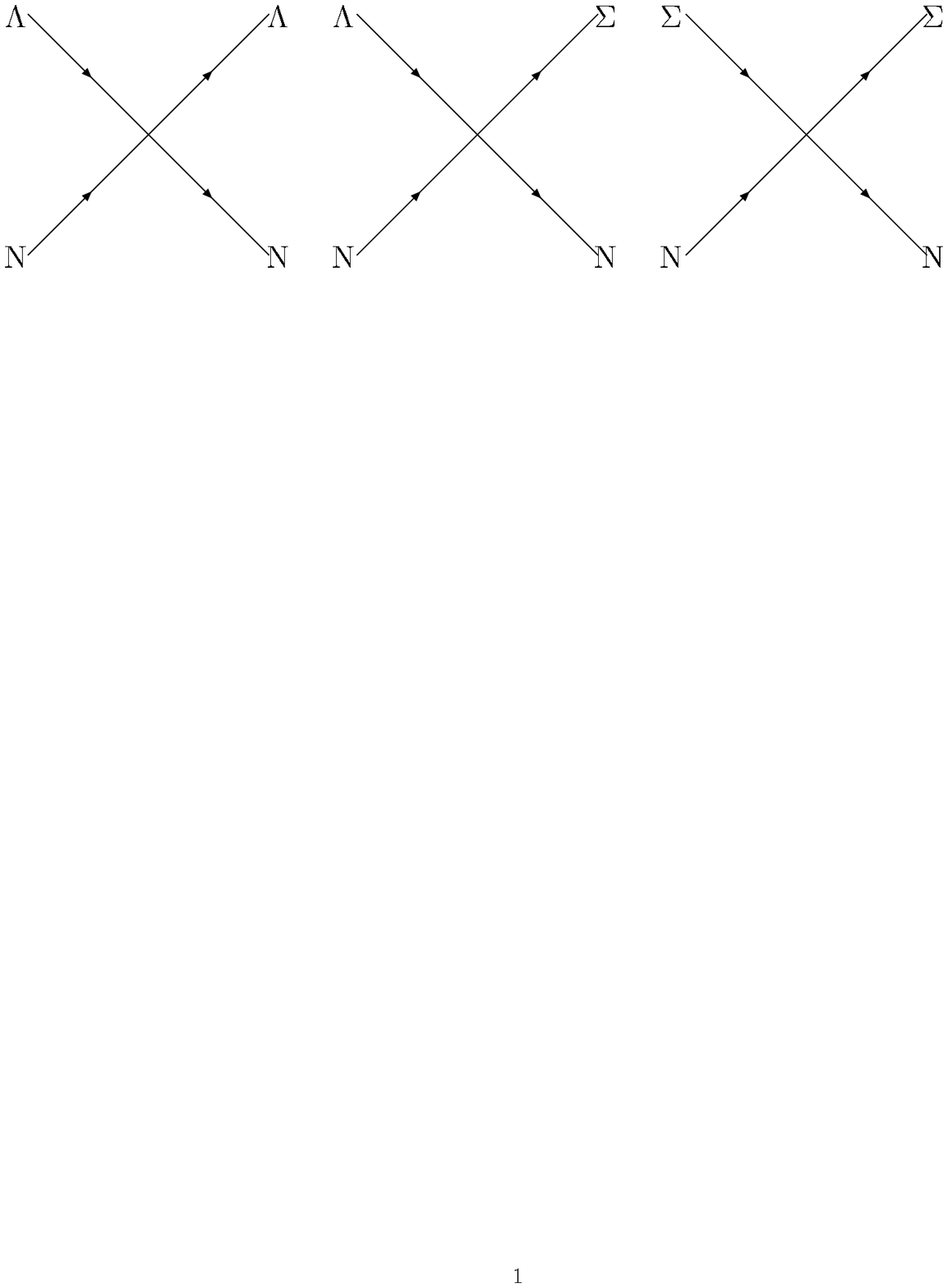}}
\end{center}
\caption{Lowest order contact terms for hyperon-nucleon interactions}
\label{fig:1}
\end{figure}

The LO contact term for the $NN$ interactions is
given by  \cite{Wei90,Epe98}
\begin{eqnarray}
{\mathcal L}&=&C_i\left(\bar{N}\Gamma_i N\right)\left(\bar{N}\Gamma_i N\right)\ ,
\label{eq:3}
\end{eqnarray}
where $\Gamma_i$ are the usual elements of the Clifford algebra \cite{Bjo65}
\begin{equation}
\Gamma_1=1 \, , \,\, 
\Gamma_2=\gamma^\mu \, , \,\,   
\Gamma_3=\sigma^{\mu\nu} \, , \,\,  
\Gamma_4=\gamma^\mu\gamma_5  \, , \,\, 
\Gamma_5=\gamma_5 \,\, . 
\label{eq:4}
\end{equation}
Considering the large components of the nucleon spinors only, the LO contact term, eq. (\ref{eq:3}), becomes
\begin{eqnarray}
{\mathcal L}
&\equiv&-\frac{C_S}{2}\left(\varphi^\dagger\varphi\right)\left(\varphi^\dagger\varphi\right)-\frac{C_T}{2}\left(\varphi^\dagger \mbox{\boldmath $\sigma$} \varphi\right)\left(\varphi^\dagger \mbox{\boldmath $\sigma$} \varphi\right)\ ,
\label{eq:5}
\end{eqnarray}
where $\varphi$ are the large components of the Dirac spinor and $C_S$ and $C_T$ are constants that need to be determined by fitting to the experimental data. In the case of $YN$ interactions we will consider a similar but SU(3)-invariant coupling. The LO contact terms for the octet baryon-baryon interactions, that are Hermitian and invariant under Lorentz transformations, are given by the SU(3) invariants,
\begin{eqnarray}
{\mathcal L}^1 &=& C^1_i \left<\bar{B}_a\bar{B}_b\left(\Gamma_i B\right)_b\left(\Gamma_i B\right)_a\right>\ , \nonumber \\
{\mathcal L}^2 &=& C^2_i \left<\bar{B}_a\left(\Gamma_i B\right)_a\bar{B}_b\left(\Gamma_i B\right)_b\right>\ , \nonumber \\
{\mathcal L}^3 &=& C^3_i \left<\bar{B}_a\left(\Gamma_i B\right)_a\right>\left<\bar{B}_b\left(\Gamma_i B\right)_b\right>\  .
\label{eq:6}
\end{eqnarray}
Here $a$ and $b$ denote the Dirac indices of the particles, $B$ is the usual irreducible octet representation of SU(3) given by
\begin{eqnarray}
B&=&
\left(
\begin{array}{ccc}
\frac{\Sigma^0}{\sqrt{2}}+\frac{\Lambda}{\sqrt{6}} & \Sigma^+ & p \\
\Sigma^- & \frac{-\Sigma^0}{\sqrt{2}}+\frac{\Lambda}{\sqrt{6}} & n \\
-\Xi^- & \Xi^0 & -\frac{2\Lambda}{\sqrt{6}}
\end{array}
\right) \ ,
\label{eq:7}
\end{eqnarray}
and the brackets denote taking the trace in the three-dimensional flavor space. The Clifford algebra elements are here actually diagonal $3\times 3$-matrices. The LO $YN$ contact terms given by these interactions are shown diagrammatically in fig. \ref{fig:1}.
The partial wave potentials are, up to an overall factor of $4\pi$,
\begin{eqnarray}
V^{NN}_{1S0}&=&2\left(C^2_S-3C^2_T\right)+2\left(C^3_S-3C^3_T\right) , \nonumber \\
V^{NN}_{3S1}&=&2\left(C^2_S+C^2_T\right)+2\left(C^3_S+C^3_T\right) , \nonumber \\
V^{\Lambda\Lambda}_{1S0}&=&\frac{1}{6}\left(C^1_S-3C^1_T\right)+\frac{5}{3}\left(C^2_S-3C^2_T\right)+2\left(C^3_S-3C^3_T\right) , \nonumber \\
V^{\Lambda\Lambda}_{3S1}&=&\frac{3}{2}\left(C^1_S+C^1_T\right)+\left(C^2_S+C^2_T\right)+2\left(C^3_S+C^3_T\right) , \nonumber \\
V^{\Sigma\Sigma}_{1S0}&=&2\left(C^2_S-3C^2_T\right)+2\left(C^3_S-3C^3_T\right) , \nonumber \\
V^{\Sigma\Sigma}_{3S1}&=&-2\left(C^2_S+C^2_T\right)+2\left(C^3_S+C^3_T\right) , \nonumber \\
\widetilde{V}^{\Sigma\Sigma}_{1S0}&=&\frac{3}{2}\left(C^1_S-3C^1_T\right)-\left(C^2_S-3C^2_T\right)+2\left(C^3_S-3C^3_T\right) , \nonumber \\
\widetilde{V}^{\Sigma\Sigma}_{3S1}&=&\frac{3}{2}\left(C^1_S+C^1_T\right)+\left(C^2_S+C^2_T\right)+2\left(C^3_S+C^3_T\right) , \nonumber \\
V^{\Lambda\Sigma}_{1S0}&=&\frac{1}{2}\left(C^1_S-3C^1_T\right)-\left(C^2_S-3C^2_T\right) , \nonumber \\
V^{\Lambda\Sigma}_{3S1}&=&-\frac{3}{2}\left(C^1_S+C^1_T\right)+\left(C^2_S+C^2_T\right) .
\label{eq:12}
\end{eqnarray}
for the $NN\rightarrow NN$, $\Lambda N\rightarrow \Lambda N$, $\Sigma N\rightarrow \Sigma N$ (I=3/2), $\Sigma N\rightarrow \Sigma N$ (I=1/2) and $\Lambda N\rightarrow \Sigma N$ channels respectively.

The six contact terms, $C_S^1$, $C_T^1$, $C_S^2$, $C_T^2$, $C_S^3$, $C_T^3$, are defined analogously to eq. \ref{eq:5} and enter the $NN$ and $YN$ potentials in only 5 different combinations. The sixth combination occurs only in the $\Lambda\Lambda$, $\Xi N$ and $\Sigma\Sigma$ channels. This is equivalent to the fact that only 5 of the $\{8\}\times\{8\}=\{27\}+\{10\}+\{10^*\}+\{8\}_s+\{8\}_a+\{1\}$ representations are relevant for $NN$ and $YN$ interactions. So, 5 contact terms need to be determined by a fit to the experimental data. Since the $NN$ data can not be described well with a LO EFT, see \cite{Wei90,Epe00a}, we will not consider the $NN$ interaction explicitly. Therefore, we consider the $YN$ partial wave potentials
\begin{eqnarray}
&&
\begin{array}{lcllcl}
V^{\Lambda\Lambda}_{1S0}&=&C^{\Lambda\Lambda}_{1S0},& V^{\Lambda\Lambda}_{3S1}&=& C^{\Lambda\Lambda}_{3S1},\\
&&&&&\\
V^{\Sigma\Sigma}_{1S0}&=&C^{\Sigma\Sigma}_{1S0},& V^{\Sigma\Sigma}_{3S1}&=&C^{\Sigma\Sigma}_{3S1},\\
&&&&&\\
\widetilde{V}^{\Sigma\Sigma}_{1S0}&=&9C^{\Lambda\Lambda}_{1S0}-8C^{\Sigma\Sigma}_{1S0},& \widetilde{V}^{\Sigma\Sigma}_{3S1}&=&C^{\Lambda\Lambda}_{3S1},\\
&&&&&\\
V^{\Lambda\Sigma}_{1S0}&=&3\left(C^{\Lambda\Lambda}_{1S0}-C^{\Sigma\Sigma}_{1S0}\right),& V^{\Lambda\Sigma}_{3S1}&=&C^{\Lambda\Sigma}_{3S1}.
\end{array}
\label{eq:13}
\end{eqnarray}
We have chosen to search for $C^{\Lambda \Lambda}_{1S0}$, $C^{\Lambda \Lambda}_{3S1}$, $C^{\Sigma \Sigma}_{1S0}$, $C^{\Sigma \Sigma}_{1S0}$, and $C^{\Lambda \Sigma}_{3S1}$ in the fitting procedure. The other partial wave potentials are then fixed by SU(3) symmetry.

\subsection{One pseudoscalar-meson exchange}
\label{sec:2.3}
Since they are discussed extensively in the literature, we show only briefly the results for one-pseudoscalar-meson exchanges. The spin-space part of the LO one-pseudoscalar-meson-exchange potential is, similar to the static one-pion-exchange potential in \cite{Epe98} (recoil and relativistic corrections give higher order contributions),
\begin{eqnarray}
V&=&-f_{B_1B_1'P}f_{B_2B_2'P}\frac{\left(\mbox{\boldmath $\sigma$}_1\cdot{\bf k}\right)\left(\mbox{\boldmath $\sigma$}_2\cdot{\bf k}\right)}{{\bf k}^2+m^2_P}\ ,
\label{eq:14}
\end{eqnarray}
where $m_P$ is the mass of the exchanged pseudoscalar meson. We defined the transferred and average momentum, ${\bf k}$ and ${\bf q}$, in terms of the final and initial center-of-mass (c.m.) momenta of the baryons, ${\bf p}'$ and ${\bf p}$, as ${\bf k}={\bf p}'-{\bf p}$ and ${\bf q}=({\bf p}'+{\bf p})/2$. The isospin part of the interaction and the SU(3)-invariant couplings $f_{BBP}$ are given e.g. in \cite{Rij99}. The physical $\eta$ was identified with the octet $\eta$ ($\eta_8$) and its physical mass was used.

\section{Scattering equation and observables}
\label{sec:3}
The calculations are done in momentum space, the scattering equation we solve for the $YN$ system is the (nonrelativistic) Lippmann-Schwinger equation. 
The Lippmann-Schwinger equation is solved in the particle basis, in order to incorporate the correct physical thresholds and the Coulomb interaction in the charged channels. The Coulomb interaction is taken into account according to the
method originally introduced by Vincent and Phatak \cite{Vin74} (see also \cite{Walzl:2000cx}). We have used relativistic kinematics for relating the laboratory energy $T_{{\rm lab}}$ of the hyperons to the c.m. momentum. Although we solve the Lippmann-Schwinger equation in the particle basis, the strong potential is calculated in the isospin basis. The potential in the Lippmann-Schwinger equation is cut off with the regulator function,
\begin{equation}
f^\Lambda(p',p)=e^{-\left(p'^4+p^4\right)/\Lambda^4}\ ,
\label{eq:15}
\end{equation}
in order to remove high-energy components of the baryon and pseudoscalar meson fields. The total cross sections are found by simply integrating the differential cross sections, except for the $\Sigma^+ p\to \Sigma^+ p$ and $\Sigma^- p\rightarrow \Sigma^- p$ channels. For those channels the experimental total cross sections were obtained via \cite{Eis71}
\begin{eqnarray}\label{eq:16}
\sigma&=&\frac{2}{\cos \theta_{{\rm max}}-\cos \theta_{{\rm min}}}\int_{\cos \theta_{{\rm min}}}^{\cos \theta_{{\rm max}}}\frac{d\sigma(\theta)}{d\cos \theta}d\cos \theta \ ,
\end{eqnarray}
for various values of $\cos \theta_{{\rm min}}$ and $\cos \theta_{{\rm max}}$. Following \cite{Rij99}, we use $\cos \theta_{{\rm min}}=-0.5$ and $\cos \theta_{{\rm max}}=0.5$ in our calculations for the $\Sigma^+ p\rightarrow \Sigma^+ p$ and $\Sigma^- p\rightarrow \Sigma^- p$ cross sections, in order to stay as close as possible to the experimental procedure.

\section{Results and discussion}
\label{sec:4}

\begin{figure}
\resizebox{0.50\textwidth}{!}{\includegraphics*[2.0cm,6.8cm][19.5cm,27cm]{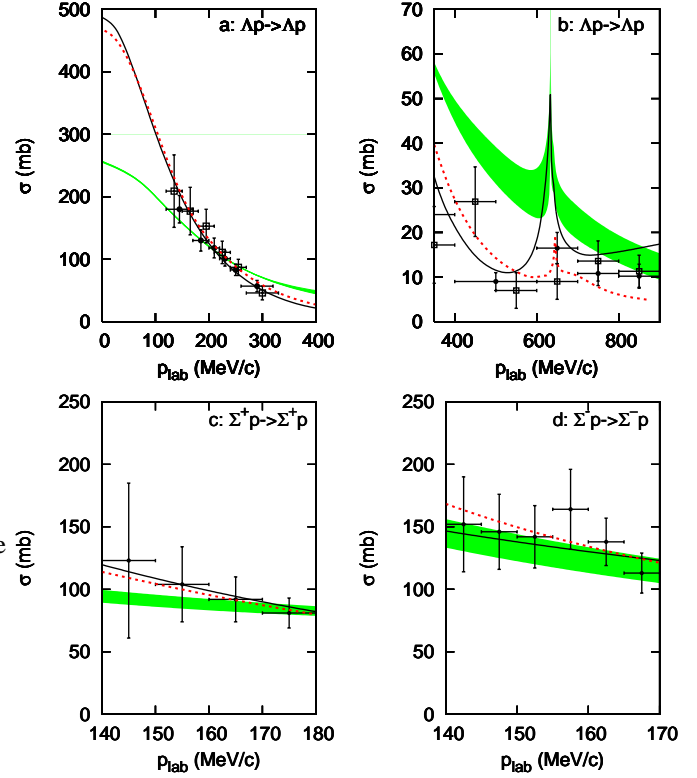}}
\caption{
''Total'' cross section $\sigma$ (as defined in eq.~(\ref{eq:16})) as a function of $p_{{\rm lab}}$. The experimental cross sections in $a$ are taken from refs.~\cite{Sec68} (open squares) and ~\cite{Ale68} (filled circles), in $b$ from refs.~\cite{Kad71} (filled circles) and \cite{Hau77} (open squares) and in $c$,$d$ from \cite{Eis71}. The shaded band is the J{\"u}lich chiral EFT for $\Lambda = 550,...,700$ MeV, the dashed curve is the J{\"u}lich '04 model \cite{Hai05}, and the solid curve is the Nijmegen NSC97f model \cite{Rij99}.
}
\label{fig:2}
\end{figure}
\begin{figure}[h]
\begin{center}
\resizebox{0.50\textwidth}{!}{\includegraphics*[2cm,6.8cm][19.5cm,27cm]{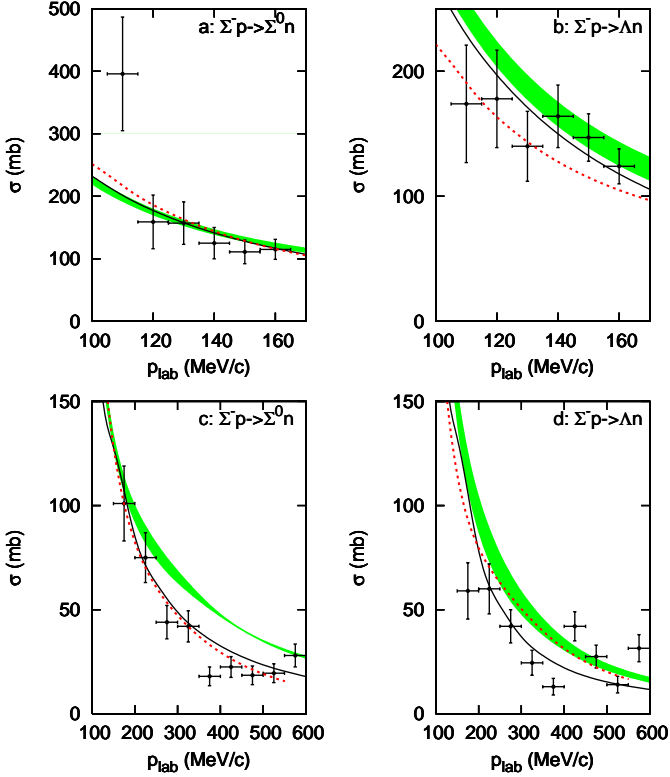}}
\end{center}
\caption{As in fig. \ref{fig:2}, but now the experimental cross sections in $a$,$b$
are taken from refs.~\cite{Eng66} and in $c$,$d$ from \cite{Ste70}.}
\label{fig:3}
\end{figure}
\begin{figure}[h]
\begin{center}
\resizebox{0.50\textwidth}{!}{\includegraphics*[2cm,6.8cm][19.5cm,27cm]{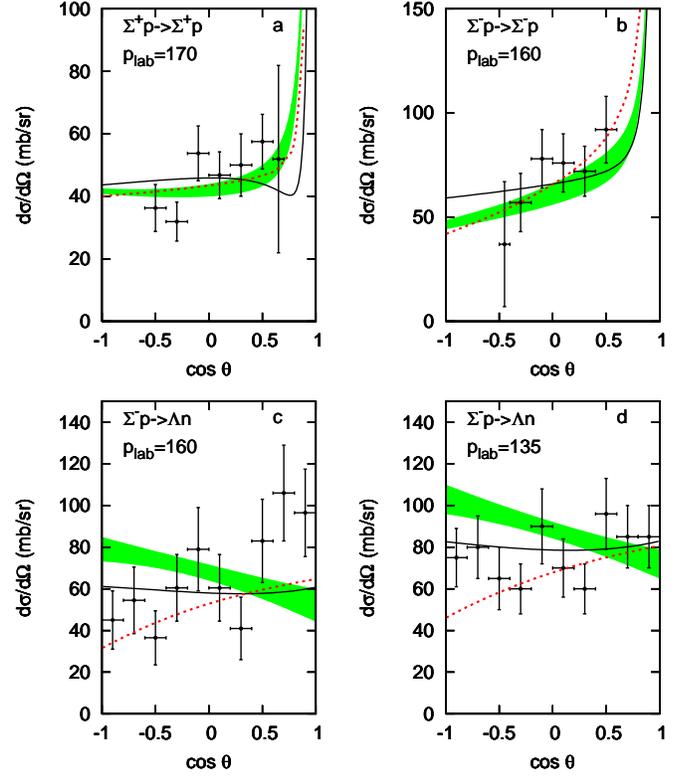}}
\end{center}
\caption{Differential cross section $d\sigma / d\cos \theta$ as a function of $\cos \theta$, where $\theta$ is the c.m. scattering angle, at various values of $p_{{\rm lab}}$ (MeV/c). The experimental differential cross sections in $a$,$b$ are taken from \cite{Eis71} and in $c$,$d$ from \cite{Eng66}. Same description of curves as in fig. \ref{fig:2}.}
\label{fig:4}
\end{figure}
\begin{figure}[h]
\begin{center}
\resizebox{0.50\textwidth}{!}{\includegraphics*[2cm,6.8cm][19.5cm,27cm]{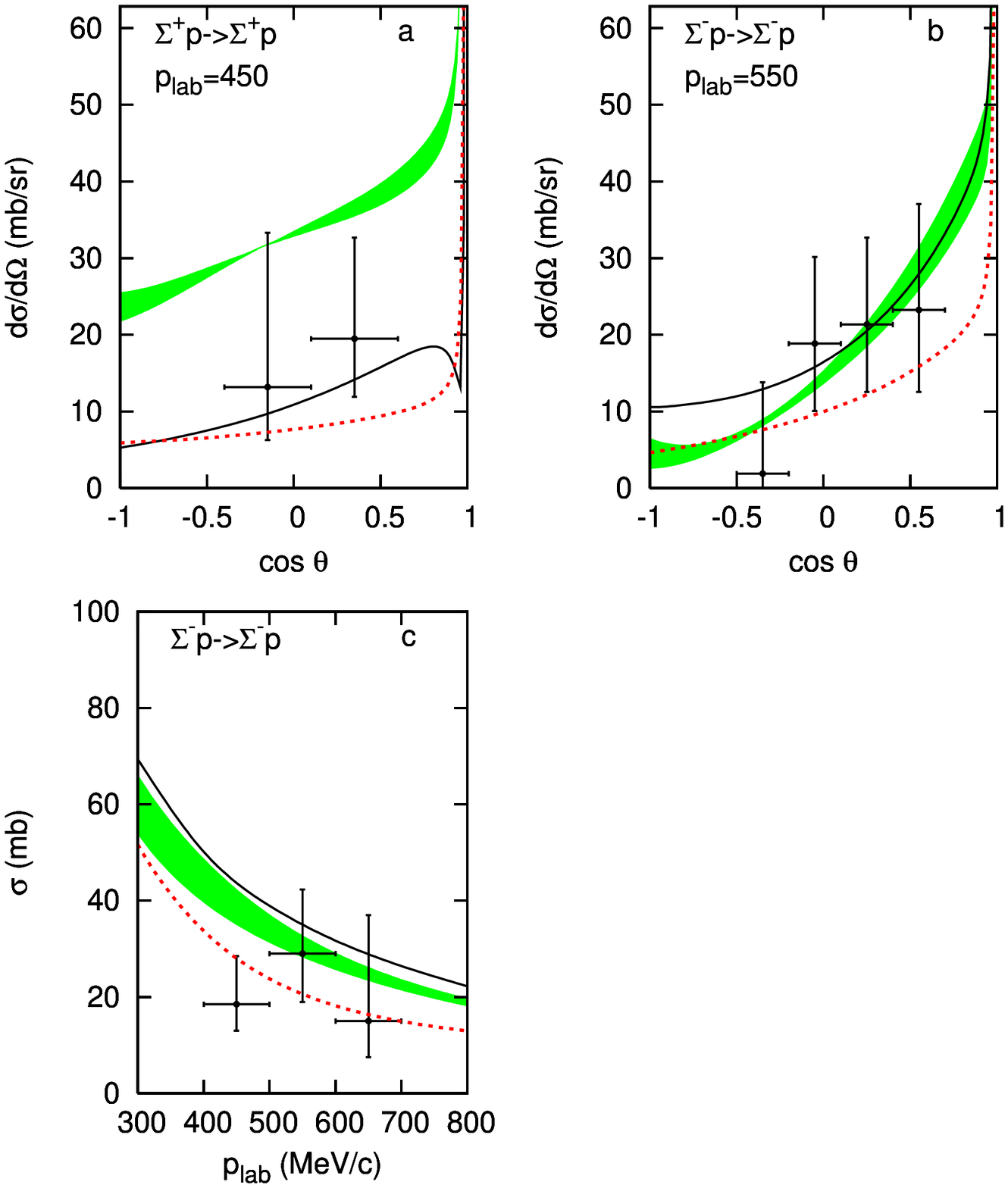}}
\end{center}
\caption{Recent $YN$ data. $a$ and $b$: differential cross section $d\sigma / d\cos \theta$ as a function of $\cos \theta$, where $\theta$ is the c.m. scattering angle, at various values of $p_{{\rm lab}}$ (MeV/c). The experimental differential cross sections are from \cite{Ahn99} and \cite{Kon00}, respectively. $c$: ''total'' cross section $\sigma$ as a function of $p_{{\rm lab}}$. The experimental cross sections are from \cite{Kon00}. Same description of curves as in fig. \ref{fig:2}.}
\label{fig:5}
\end{figure}

For the fitting procedure we consider the empirical low-energy total cross sections shown in figs. \ref{fig:2}a,c, and d and \ref{fig:3}a and b, and the inelastic capture ratio at rest \cite{Swa62}, in total 35 $YN$ data. These data have also been used in \cite{Hai05,Rij99}. The higher energy total cross sections and differential cross sections are then predictions of the LO chiral EFT, which contains five free parameters. The fits are done for fixed values of the cut-off mass ($\Lambda$) and of $\alpha$, the pseudoscalar $F/(F+D)$ ratio.

The five LECs in eq. (\ref{eq:13}), were varied during the parameter search to the set of 35 low-energy $YN$ data. The values obtained in the fitting procedure for cut-off values between $550$ and $700$ MeV, are listed in table \ref{tab:1}. We remark that the range of cut-off values is similar to the range in the $NN$ case, e.g. \cite{Epe02,Epe04}. The range is limited from below by the mass of the pseudoscalar mesons. Since we do a LO calculation we do not expect a large plateau (i.e. a practically stable $\chi^2$ for varying $\Lambda$). Note that recently the cut-off dependence of EFT calculations has been discussed extensively in the literature, see e.g. \cite{Nog05,Val04,Epe06}.
\begin{table}[t]
\caption{The $YN$ $S$-wave contact terms for various cut-offs. The LECs are in $10^4$ ${\rm GeV}^{-2}$; $\Lambda$ is in MeV. $\chi^2$ is the total chi squared for 35 $YN$ data.}
\label{tab:1}
\centering
\begin{tabular}{rrrrr}
\hline\noalign{\smallskip}
$\Lambda$& $550$& $600$& $650$& $700$  \\
\noalign{\smallskip}\hline\noalign{\smallskip}
$C^{\Lambda \Lambda}_{1S0}$ &$-.0466$ &$-.0403$ &$-.0322$ &$-.0304$\\
$C^{\Lambda \Lambda}_{3S1}$ &$-.0222$ &$-.0163$ &$-.0097$ &$-.0022$\\
$C^{\Sigma \Sigma}_{1S0}$   &$-.0766$ &$-.0763$ &$-.0757$ &$-.0744$\\
$C^{\Sigma \Sigma}_{3S1}$    &$.2336$  &$.2391$  &$.2392$  &$.2501$\\
$C^{\Lambda \Sigma}_{3S1}$  &$-.0016$ &$-.0019$  &$.0000$  &$.0035$\\
\noalign{\smallskip}\hline\noalign{\smallskip}
$\chi^2$& 29.6& 28.3& 30.3& 34.6\\
\noalign{\smallskip}\hline
\end{tabular}
\end{table}

The fits were first done for the cut-off mass $\Lambda =600$ MeV. We remark that the $\Lambda N$ $S$-wave scattering lengths resulting for that cut-off were then kept fixed in the subsequent fits for the other cut-off values. We did this because the $\Lambda N$ scattering lengths are not well determined by the scattering data. As a matter of fact, not even the relative magnitude of the $\Lambda N$ triplet and singlet interaction can be constrained from the $YN$ data, but their strengths play an important role for the hypertriton binding energy \cite{YNN1}. Contrary to the $NN$ case, see, e.g. \cite{Epe00a}, the contact terms are in general not determined by a specific phase shift, because of the coupled particle channels in the $YN$ interaction. Furthermore, due to the limited accuracy and incompleteness of the $YN$ scattering data there are no unique partial wave analyses. Therefore we have fitted the chiral EFT directly to the cross sections. A good description of the considered $YN$ scattering data has been obtained in the considered cut-off region, as can be seen in table \ref{tab:1} and figs. \ref{fig:2}a,c,d and \ref{fig:3}a,b. In these figures the shaded band represents the results of the chiral EFT in the considered cut-off region. In this low-energy regime the cross sections are mainly given by the $S$-wave contribution, except for for the $\Lambda N\rightarrow \Sigma N$ cross section where the ${}^3D_1(\Lambda N)\leftrightarrow {}^3S_1(\Sigma N)$ transition provides the main contribution. Still all partial waves with total angular momentum $J\le 2$ were included in the computation of the observables. The $\Lambda p$ cross section shows a clear cusp at the $\Sigma^+ n$ threshold, peaking at 60 mb. Fig. \ref{fig:2}b shows that the predicted $\Lambda p$ cross section at higher energies is too large, which is related to the problem that some LO phase shifts are too large at higher energies. Note that this was also the case for the $NN$ interaction \cite{Epe00a}. In a NLO calculation this problem will probably vanish. The differential cross sections at low energies, which have not been taken into account in the fitting procedure, are predicted well, see fig. \ref{fig:4}. The results of the chiral EFT are also in good agreement with the scattering data at higher energy, the older ones in figs. \ref{fig:3}c,d as well as the
more recent scattering data in fig. \ref{fig:5}.

We have, so far, used the SU(6) value for the pseudoscalar $F/(F+D)$ ratio; $\alpha=0.4$. We studied the dependence on this parameter by varying it within a range of 10 percent. After refitting the contact terms we basically found an equally good description of the empirical data. Therefore, we keep $\alpha$ to its SU(6) value. An uncertainty in our calculation is the value of the $\eta$ coupling, since we identified the physical $\eta$ with the octet $\eta$. Therefore, we varied the $\eta$ coupling between zero and its octet value, but we found very little influence on the description of the data (in fact, inclusion of the $\eta$ leads to a better plateau in the cut-off range considered). Also baryon mass differences squared in the propagator in eq. (\ref{eq:14}) are not consistent with our power counting and were not considered, but we studied their influence on the description of the data. We found that the quality of the description does not depend on these terms.

\begin{figure}[t]
\resizebox{0.50\textwidth}{!}{\includegraphics*[2.0cm,6.8cm][19.5cm,27cm]{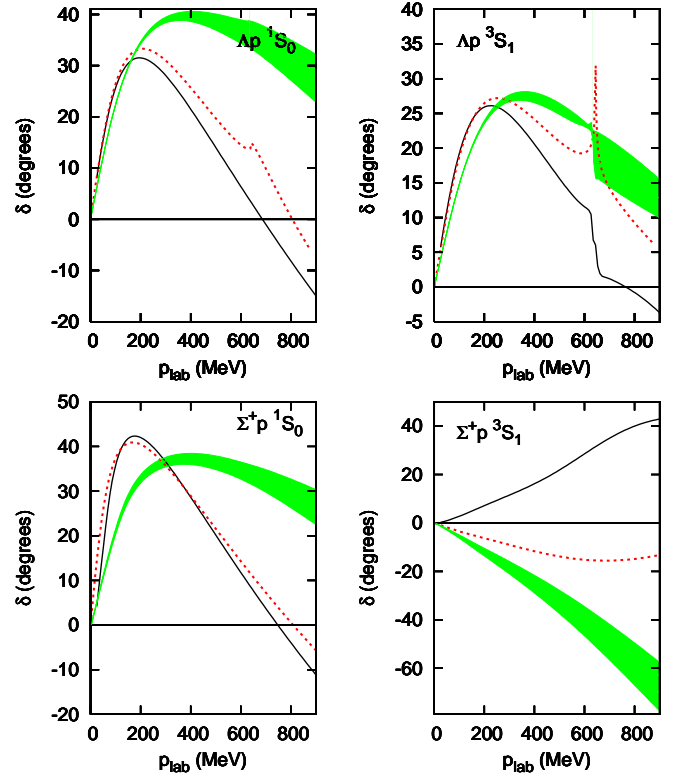}}
\caption{The $\Lambda p$ and $\Sigma^+ p$ $S$-wave phase shifts $\delta$ as a function of $p_{{\rm lab}}$. Same description of curves as in fig. \ref{fig:2}.}
\label{fig:6}
\end{figure}

The $S$-wave phase shifts for $\Lambda p$ and $\Sigma^+p$ are shown in fig. \ref{fig:6}. 
As mentioned before, the limited accuracy of the $YN$ scattering data does not allow for a unique phase shift analysis. This explains why the chiral EFT phase shifts are quite different from the phase shifts of the models presented in refs.~\cite{Hai05,Rij99}. The predictions of the latter models also differ between each other in many partial waves. In both the $\Lambda p$ and $\Sigma^+ p$ ${}^1S_0$ partial waves, the LO chiral EFT phase shifts are much larger at higher energies than the phases from \cite{Hai05,Rij99}. We emphasize that the empirical data, considered in the fitting procedure, are at lower energies. Also for the $NN$ interaction in leading order these partial waves were much larger than the Nijmegen phase shift analysis, see \cite{Epe00a}. It is expected that this problem for the $YN$ interaction can be solved by the derivative contact terms in a NLO calculation, just like in the $NN$ case. Our ${}^3S_1$ $\Sigma^+ p$ phase shift is repulsive like in \cite{Hai05}, but contrary to \cite{Rij99}.

The $\Lambda p$ and $\Sigma^+p$ scattering lengths are listed in table \ref{tab:2} together with the corresponding hypertriton binding energies (preliminary results of $YNN$ Faddeev calculations from \cite{Nog06}).
\begin{table}[b]
\caption{The predicted $YN$ singlet and triplet scattering lengths (in fm) and hypertriton binding energy, $E_B$ (in MeV). Experimentally the deuteron binding energy is $-2.224$~MeV and the hypertriton binding energy is $-2.354(50)$~MeV. The binding energies for the hypertriton, \cite{Nog06}, are calculated using the Idaho-N3LO $NN$ potential \cite{Entem:2003ft}.}
\label{tab:2}
\centering
\begin{tabular}{rrrrr}
\hline\noalign{\smallskip}
$\Lambda$& 550& 600& 650& 700  \\
\noalign{\smallskip}\hline\noalign{\smallskip}
$a^{\Lambda p}_s$ &$-1.90$ &$-1.91$ &$-1.91$ &$-1.91$ \\
$a^{\Lambda p}_t$ &$-1.22$ &$-1.23$ &$-1.23$ &$-1.23$ \\
\noalign{\smallskip}\hline\noalign{\smallskip}
$a^{\Sigma^+ p}_s$  &$-2.24$ &$-2.32$  &$-2.36$  &$-2.29$ \\
$a^{\Sigma^+ p}_t$   &$0.70$  &$0.65$   &$0.60$   &$0.56$ \\
\noalign{\smallskip}\hline\noalign{\smallskip}
$E_B$ &$-2.35$ &$-2.34$ &$-2.34$ &$-2.36$  \\
\noalign{\smallskip}\hline
\end{tabular}
\end{table}
Our singlet $\Sigma^+ p$ scattering length is about half as large as the values found in \cite{Hai05,Rij99}. Similar to those models and other $YN$ interactions, the value of the triplet $\Sigma^+ p$ scattering length is rather small. Contrary to \cite{Rij99}, but similar to \cite{Hai05} we found repulsion in this partial wave. The magnitudes of the $\Lambda p$ scattering lengths are smaller than the corresponding values of the Nijmegen NSC97f and J{\"u}lich '04 models \cite{Hai05,Rij99}, which is also reflected in the small $\Lambda p$ cross section near threshold, see fig. \ref{fig:2}a. The mentioned models lead to a bound hypertriton \cite{Nog06,YNN2}. Although our $\Lambda p$ scattering lengths differ significantly from those of \cite{Hai05,Rij99}, the $YN$ interaction based on chiral EFT also yields a correctly bound hypertriton, see table \ref{tab:2}. Preliminary results for the four-body hypernuclei ${}^4_\Lambda {\rm H}$ and ${}^4_\Lambda {\rm He}$, see \cite{Nog06a}, show that the chiral EFT predicts reasonable $\Lambda$ separation energies for ${}^4_\Lambda {\rm H}$, but the charge dependence of the $\Lambda$ separation energies is not reproduced.

Finally, our findings show that the chiral EFT scheme, applied in ref.~\cite{Epe05} to the $NN$ interaction, also works well for the $YN$ interaction. In the future it will be interesting to perform a combined $NN$ and $YN$ study in chiral EFT, starting with a NLO calculation. Also an SU(3) extension to the hyperon-hyperon ($YY$) sector is of interest. Work in this direction is in progress.

\begin{acknowledgement}
I thank Johann Haidenbaur and Ulf-G. Mei\ss ner for collaborating on this work. Also I am very grateful to Andreas Nogga for providing me with the hypernuclei results.
\end{acknowledgement}

\bibliographystyle{phaip}
\bibliography{polinder}

\end{document}